\documentclass[12pt,preprint]{aastex}

\slugcomment{accepted for publication in ApJ}

\shorttitle{Bias and Conditional Mass Function of Dark Halos Based
on the Nonspherical Collapse Model} \shortauthors{Lin, Chiueh /&
Lee}

\begin{document}

\title{Bias and Conditional Mass Function of Dark Halos Based
on the Nonspherical Collapse Model}

\author{Lihwai Lin\altaffilmark{1}, Tzihong Chiueh\altaffilmark{2}}
\affil{Department of Physics, National Taiwan University, Taipei,
Taiwan}

\and

\author{Jounghun Lee\altaffilmark{3}}
\affil{Institute of Astronomy and Astrophysics, Academia Sinica,
Taipei, Taiwan}

\altaffiltext{1}{Email: d90222005@ms90.ntu.edu.tw}
\altaffiltext{2}{Email: chiuehth@phys.ntu.edu.tw}
\altaffiltext{3}{Email: taiji@asiaa.sinica.edu.tw}

\begin{abstract}
Nonspherical collapse is modelled, under the Zeldovich
approximation, by six-dimensional random walks of the initial
deformation tensor field. The collapse boundary adopted here is a
slightly-modified version of that proposed by Chiueh and Lee
(2001). Not only the mass function agrees with the fitting formula
of Sheth and Tormen (1999), but the bias function and conditional
mass function constructed by this model are also found to agree
reasonably well with the simulation results of Jing (1998) and
Somerville et al. (2000), respectively. In particular, by
introducing a small mass gap, we find a fitting formula for the
conditional mass function, which works well even at small time
intervals between parent and progenitor halos during the merging
history.
\end{abstract}


\section{Introduction}

In a hierarchical model, the gravitationally bound objects of the
universe formed from mergers of smaller objects into larger ones.
An issue of fundamental importance is to understand how
gravitational structures, such as galaxies and clusters, are
distributed as a function of mass and evolving time. Press \&
Schechter (1974) provided a simple analytical model (hereafter PS
model) to describe the mass distribution of dark halos (mass
function). This theory is based on the assumption that the process
of gravitational collapse can be approximated by spherical
symmetry and it occurs when the mean density contrast of
precollapse regions predicted by the linear theory reaches a
threshold value $\delta_{c}$. Bond et al. (1991) introduced the
excursion set formalism and extended the PS theory to calculations
of the conditional mass function, which is closely related to the
issue of the linear bias and halo merging histories. Mo \& White
(1996) used the extended Press-Schechter theory (hereafter EPS) to
address the large-scale bias, which concerns how the collapsed
halos trace the underlying dark matter. Moreover, the EPS theory
is also useful for construction of the merger history of dark
halos (Bower 1991; Lacey \& Cole 1993, 1994 hereafter LC93, LC94),
which concerns the probability that a halo of mass $M_{2}$ at time
$t_{2}$ has a progenitor halo of mass $M_{1}$ at earlier time
$t_{1}$.

The PS or EPS theory has been tested against N-body simulations
(Efstathiou et al. 1988; LC94), and the comparisons suggest that
the PS model overestimates the number of halos in low-mass regimes
and underestimates it in high-mass regimes. This discrepancy
indicates that the spherical collapse model used by the PS theory
may be too simplistic. Many researchers have taken the
nonspherical collapse process into account to correct this problem
(e.g., Bond \& Myers 1996; Monaco 1995; Lee \& Shandarin 1998;
Chiueh \& Lee 2001; Sheth, Mo and Tormen 2001). Most of these
works are still based on the main idea of the PS theory, and
further incorporate either the dynamics of ellipsoids or the
Zeldovich approximation (Zeldovich 1970).

The main aims of this paper are to investigate the large-scale
bias and conditional mass function by using the nonspherical
collapse boundary (hereafter NCB) proposed by Chiueh and Lee
(2001). We also test whether the results derived from nonspherical
collapse boundary can provide better statistical predictions of
dark halos than the EPS model does.

The outline of the paper is as follows. In $\S$ 2, we summerize
the PS and EPS theory, and briefly review the Zeldovich
approximation and the nonspherical collapse boundary proposed by
Chiueh \& Lee (2001). In $\S$ 3, we show the algorithm and the
results of mass function and large-scale bias of our model. In
$\S$ 4, the conditional mass function is derived and the
comparison with N-body simulation is presented. In $\S$ 5, the
conclusion is given.

\section{(Extended) Press-Schechter Formalism}

\subsection{Mass function}The Press-Schechter theory
(1974) assumes that the initial density fluctuations are Gaussian
random fields. Let $\delta_{R}(\vec{x})$ denote the overdensity
field at position $\vec{x}$ in the comoving coordinate when the
density distribution is smoothed over scale R by a window function
$W_{R}$. The relation is
\begin{equation}
 \delta_{R}(\vec{x})=\int\delta(\vec{x}')W_{R}(\vec{x}-\vec{x}')d^{3}\vec{x}',
\end{equation}
with a squared variance
\begin{equation}\label{eq2}
 S=\sigma^{2}(M)=<|\delta_{R}(\vec{x})|^{2}>=\frac{1}{(2\pi)^{3}} \int p(k)\widetilde{W}_{R}^{2}(k)4
 \pi k^{2}dk,
\end{equation}
where $p(k)$ is the power spectrum  of density fluctuations and
$\widetilde{W}_{R}(k)$ is the Fourier transform of the window
function $W_{R}(\vec{x}-\vec{x}')$. The squared variance $S$
increases as mass $M$ decreases, and in this paper, we will often
use $S$ to represent the mass scale.

Bond et al. (1991) extended the original PS method and used random
walks (or called excursion set) to address the mass function
problem. When variance $S$ equals zero, the density contrast
$\delta$ is set to zero. As $S$ increases, $\delta$ changes
randomly. When $\delta$ first hits the threshold value
$\delta_{c}$, the random walk stops and the region of mass scale
$S$, over which $\delta$ is smoothed, collapses into a bound
object. This prescription is equivalent to a diffusion process
with an absorbing boundary at $\delta=\delta_{c}$, which has been
solved by Chandrasekhar (1943) when the window function is chosen
to be a sharp k-space filter. Thus, one can calculate the
probability of a trajectory having its first hit at the threshold
$\delta_{c}$ in the mass range between $S$ and $S+dS$ to represent
the mass function, which is described by
\begin{equation}
f(S,\delta_{c})dS=\frac{\delta_{c}}{\sqrt{2\pi}S^{3/2}}\mathrm{exp}(-\frac{\delta_{c}^{2}}{2S})dS,
\end{equation}
where the functional form $f(x)$ represents the differential
probability in the range between $x$ and $x+dx$. Moreover, the
(comoving) number density of halos with mass M is
\begin{equation}\label{eq4}
\frac{dn}{dM}dM=\frac{\rho_{b}}{M}f(S,\delta_{c})|\frac{dS}{dM}|dM,
\end{equation}
where $\rho_{b}$ is the mean mass density of the universe.
Alternative choices of window functions, such as real-space
top-hat filter and Gaussian filter, are also often used to
determine the relation of the squared variance $S$ and mass M. The
localized filters yield correlated steps in the above random-walk
framework. Comparisons among different types of window function
are well discussed by Bond et al. (1991), LC93 and LC94.

For simplicity, we define a variable $\nu\equiv\delta_{c}/S^{1/2}$
and one can easily show that the mass function can be represented
as
\begin{equation}
\nu f(\nu)d\nu=\sqrt{\frac{2}{\pi}}\nu
\mathrm{exp}(-\frac{\nu^{2}}{2})d\nu.
\end{equation}
Here $\delta_{c}$ is often chosen as 1.69, which is predicted by
the spherical collapse model. The main advantage of the excusion
set approach is that while the mass function depends on the
variance $S$ and the value of collapse threshold $\delta_{c}$, the
cosmological models enter this problem only through the relation
between $S$ and the mass $M$.

Upon comparing with N-body simulations, one finds that the EPS
formalism overestimates low-mass and underestimates high-mass
halos. Of course one may vary the value of the collapse threshold
$\delta_{c}$, but it is not possible to match simulation results
well for all mass ranges.

\subsection{Conditional mass function}In the picture of the excursion set theory, the conditional
probability represents the fraction of trajectories first hitting
a higher threshold $\delta_{c1}$ among those trajectories that
start from a lower threshold $\delta_{c2}$. In the framework of
EPS theory, Lacey \& Cole (1993) proposed that the conditional
probability can assume a modified form where the origin of the
mass function $(0,0)$ shifts to ($S_{2},\delta_{c2}$). Therefore,
the conditional probability is simply
\begin{equation}\label{conditionaleq}
f(S_{1},\delta_{c1}|S_{2},\delta_{c2})dS_{1}=\frac{(\delta_{c1}-\delta_{c2})}{\sqrt{2\pi}(S_{1}-S_{2})^{3/2}}\mathrm{exp}[-\frac{(\delta_{c1}-\delta_{c2})^{2}}{2(S_{1}-S_{2})}]dS_{1}.
\end{equation}
Here, it demands $S_{1}>S_{2}$ and $\delta_{c1}>\delta_{c2}$.

Conditional probability is useful for investigation of large-scale
bias as well as merger events of halos. In the first case, the
physical meaning of $\delta_{c2}$ is not the collapse threshold at
some specific time, but denotes the mean density contrast of a
large uncollapsed region of mass scale $S_{2}$. Thus we shall
change $\delta_{c2}$ into $\delta_{2}$ to avoid confusion. If we
define the bias parameter $b$ to be the fraction of excess dark
halos with mass scale $S_{1}$ that can be found in a larger mass
scale $S_{2}$, then under the limit $S_{2}\gg S_{1}$ and
$\delta_{2}\ll\delta_{c1}$, Mo \& White (1996, hereafter MW96) use
the EPS formalism to derive that
\begin{equation}\label{eqbias}
b=\frac{(\delta_{c1}/S_{1}^{1/2})^{2}-1}{\delta_{c1}}=\frac{\nu_{1}^{2}-1}{\delta_{c1}},
\end{equation}
where $\nu_{1}\equiv\delta_{c1}/S_{1}^{1/2}$. For convenience, we
define a characteristic mass $M^{*}$, which satisfies the
requirement $S(M^{*})=\delta_{c1}^{2}$. When
$S_{1}<\delta_{c1}^{2}$, i.e., $M_{1}>M^{*}$, the bias factor $b$
is positive, which means the density of halos with masses greater
than $M^{*}$ is larger in the denser region of dark matter. On the
other hand, when $S_{1}>\delta_{c1}^{2}$, or $M_{1}<M^{*}$, the
opposite holds.

In the case of merger events, $\delta_{c2}$ denotes the collapse
threshold at a later time $t_{2}$ (as compared with $t_{1}$).
Again, the meanings of notations change: $\delta_{c1} \rightarrow
\delta_{c}(z_{1}$) and $\delta_{c2} \rightarrow
\delta_{c}(z_{2})$, the collapse thresholds at redshifts $z_{1}$
and $z_{2}$, respectively. Thus,
$f(M_{1},\delta_{c1}|M_{2},\delta_{c2})dM_{1}$ represents the mass
fraction of $M_{2}$ at redshift $z_{2}$ that is in the form of
collapsed halos of mass $M_{1}$ in an earlier epoch ${z_{1}}$.

\subsection{Zeldovich approximation}Zeldovich (1970)
provided a good way to describe the development of density
perturbations into the nonlinear regime. Let $a(t)$ denote the
uniform expansion of the background. The physical coordinate
$\vec{r}$ and the comoving coordinate $\vec{x}$ are simply
connected by
\begin{equation}
\vec{r}=a(t)\vec{x}.
\end{equation}
In the Zeldovich aprroximation, the motion of a particle is
described by
\begin{equation}
\vec{x}=\vec{q}+b(t)\vec{p}(\vec{q}),
\end{equation}
where $\vec{q}$ is the Lagrangian coordinate (initial location) of
a particle. The second term is the perturbation of a particle in
the Lagrangian coordinate. From the conservation of mass
condition, we have
\begin{equation}
\rho(\vec{x},t)d^{3}\vec{x}=\bar{\rho}(\vec{q},t)d^{3}\vec{q}=\bar{\rho}(\vec{q},0)d^{3}\vec{q}.
\end{equation}
It implies
\begin{equation}\label{eq11}
\rho(\vec{x},t)=\bar{\rho}\frac{d^{3}\vec{q}}{d^{3}\vec{x}}=\bar{\rho}\frac{1}{det(\frac{d^{3}\vec{x}}{d^{3}\vec{q}})}\equiv\bar{\rho}\frac{1}{det\hat{J}}.
\end{equation}
The tensor $\hat{J}$ in the denominator is the sum of an identity
tensor and a deformation tensor $\hat{D}$. Let
$p(\vec{q})=-\nabla_{\vec{q}}\Phi(\vec{q})$,then
\begin{equation}
D_{ij}=b(t)\frac{\partial p_{i}}{\partial q_{i}}= -
b(t)\frac{\partial^{2}}{\partial q_{i}\partial q_{j}}\Phi.
\end{equation}
Because $\hat{D}$ is real and symmetric, it has three real
eigenvalues, $\Lambda= (-\lambda_{1}, -\lambda_{2},
-\lambda_{3})$, at every spatial location. Thus, Eq. (\ref{eq11})
can be rewritten as
\begin{equation}\label{eq13}
\rho(\vec{x},t)=\bar{\rho}\frac{1}{(1-\lambda_{1})(1-\lambda_{2})(1-\lambda_{3})}.
\end{equation}
When $\lambda's$ are small, the density contrast $\delta$ is
\begin{equation}\label{delta1}
\delta=\frac{\rho(\vec{x},t)-\bar{\rho}}{\bar{\rho}}\sim
(\lambda_{1}+\lambda_{2}+\lambda_{3}).
\end{equation}
The sign of the value of $\lambda$ determines the divergence
(negative $\lambda$) or convergence (positive $\lambda$) of the
local motion. Some examples illustrate the physical meaning of Eq.
(\ref{eq13}). If $\lambda_{1}=\lambda_{2}=\lambda_{3}$, when all
$\lambda$'s approach unity the density becomes infinity, and it
implies the spherical collapse case. If a certain $\lambda_{i}$
grows to unity before the other two $\lambda$'s do, then the local
ellipsoid will collapse into a pancake. It is easy to realize that
the latter be much more probable than the former, and therefore
the gravitational collapse should be nonspherical.

Since the deformation tensor $\hat{D}$ is symmetric, it has $6$
degrees of freedom: $d_{11}$, $d_{22}$, $d_{33}$, $d_{12}$,
$d_{23}$ and $d_{13}$. The first three represent the diagonal
terms, and the rest represent the off-diagonal ones. In addition,
one can construct $6$ independent Gaussian variables, namely, $Y=
(y_{1},y_{2},y_{3},y_{4},y_{5},y_{6})$. The $6$ components of the
deformation tensor can be interpreted as linear combinations of
$y_{1}\sim y_{6}$:
\begin{mathletters}\label{delta2}
\begin{equation}
d_{11}=-(y_{1}+\frac{3}{\sqrt{15}}y_{2}+\frac{1}{\sqrt{5}}y_{3})/3,
\end{equation}
\begin{equation}
d_{22}=-(y_{1}-\frac{2}{\sqrt{5}}y_{3})/3,
\end{equation}
\begin{equation}
d_{33}=-(y_{1}-\frac{3}{\sqrt{15}}y_{2}+\frac{1}{\sqrt{5}}y_{3})/3,
\end{equation}
\begin{equation}
d_{12}=\frac{1}{15}y_{4},
\end{equation}
\begin{equation}
d_{23}=\frac{1}{15}y_{5},
\end{equation}
\begin{equation}
d_{13}=\frac{1}{15}y_{6}.
\end{equation}
\end{mathletters}

\subsection{Nonspherical collapse boundary (NCB)}In
order to resolve the discrepancy of the mass functions between the
PS prediction and the N-body simulation, Chiueh \& Lee (2001)
investigate how the nonspherical collapse process may be relevant.
In addition to adopting the fundamental concepts of the PS
hypothesis and the excursion set approach, they have noticed that
the collapse condition should contain full information of the
three eigenvalues ($\lambda_{1}, \lambda_{2}, \lambda_{3}$), and
that the nonspherical effect is more dominant in the low-mass
regime than in the high-mass regime. Taking these properties into
account, they propose a nonspherical collapse boundary (NCB
hereafter) having the following form:
\begin{equation}
\frac{\delta}{\delta_{c}}=(1+\frac{r^{4}}{\beta})^{\beta},
\end{equation}
where $\beta$ is a fitting parameter taken to be 0.15,
$\delta_{c}$ remains the collapse threshold predicted by the
spherical model, which is taken to be 1.5 by Chiueh \& Lee (2001)
rather than 1.69 due to the correction given by Shapiro et al.
(1999), and the definition of $r^{2}$ is
\begin{equation}
r^{2}\equiv\frac{1}{3}[(\lambda_{1}-\lambda_{2})^{2}+(\lambda_{2}-\lambda_{3})^{2}+(\lambda_{3}-\lambda_{1})^{2}].
\end{equation}
In the case of spherical collapse, the three eigenvalues equal to
each other. Therefore, the quantity $r$ can be regarded as an
indicator for the degree of nonsphericality.

The form of NCB has several advantages. First, this boundary is
smooth and rotational invariant in the eigenvalue
($\lambda_{1},\lambda_{2},\lambda_{3}$) space. Second, when $r$ is
close to zero, the NCB is reduced to the spherical collapse
condition as expected. Third, the variable $r^{2}$ is related to
the angular momentum (Catelan \& Theuns 1996), so that the NCB
provides additional information about the nonspherical collapse
halos.

\section{Mass Function and Bias for NCB}

\subsection{Mass function}Here we slightly improve the
nonspherical collapse boundary of Chiueh and Lee (2001) into this
form:
\begin{equation}\label{boundary}
\frac{\delta}{\delta_{c}}=(1+\frac{r^{4}}{\beta_{1}})^{\beta_{2}},
\end{equation}
where $\delta_{c}=1.5654$, $\beta_{1}=0.26$ and $\beta_{2}=0.159$.
The three symbols $\delta_{c}$, $\beta_{1}$ and $\beta_{2}$ are
all fitting parameters to obtain a mass function closest to the
result of N-body simulation (Sheth \& Tormen 1999, hereafter
ST99). Fine tuning of these parameters would not change the entire
shape of mass function too much but it is necessary for us to
perform the bias problem. The eigenvalues $\lambda$ satisfy the
eigenvalue equation
\begin{eqnarray}
\lambda^{3}+(d_{11}+d_{22}+d_{33})\lambda^{2} +
(d_{11}d_{22}+d_{22}d_{33}+d_{33}d_{11}-d_{12}^{2}-d_{23}^{2}-d_{13}^{2})\lambda
- \nonumber\\
 \  \
  (d_{11}d_{23}^{2}+d_{22}d_{13}^{2}+d_{33}d_{12}^{2}-d_{11}d_{22}d_{33}-2d_{12}d_{23}d_{13})=0.
\end{eqnarray}
Therefore,
\begin{mathletters}
\begin{equation}\label{delta3}
\lambda_{1}+\lambda_{2}+\lambda_{3}=-(d_{11}+d_{22}+d_{33}),
\end{equation}
\begin{equation}\label{lambda}
\lambda_{1}\lambda_{2}+
\lambda_{2}\lambda_{3}+\lambda_{3}\lambda_{1}
=(d_{11}d_{22}+d_{22}d_{33}+d_{33}d_{11}-d_{12}^{2}-d_{23}^{2}-d_{13}^{2}),
\end{equation}
\begin{equation}
\lambda_{1}\lambda_{2}\lambda_{3}=(d_{11}d_{23}^{2}+d_{22}d_{13}^{2}+d_{33}d_{12}^{2}-d_{11}d_{22}d_{33}-2d_{12}d_{23}d_{13}).\\
\end{equation}
\end{mathletters}
Using Eqs. (\ref{delta2}), (\ref{delta3}) and (\ref{lambda}), it
is easy to convert $r^{2}$ into a function of $y_{1}, y_{2},
y_{3}, y_{4}, y_{5}$, and $y_{6}$. The result is
\begin{equation}
r^{2}=\frac{2}{15}(y_{2}^{2}+y_{3}^{2}+y_{4}^{2}+y_{5}^{2}+y_{6}^{2}).
\end{equation}
This neat equation has been pointed out by Sheth \& Tormen (2002).
Combined with $\delta=y_{1}$ (from Eqs. (\ref{delta1}),
(\ref{delta2}) and (\ref{delta3})), we can simulate the random
process in the 6-dimensional $Y$-space with an absorbing boundary,
instead of conducting the equivalent, but more complicated, random
walks in $(\lambda_{1}, \lambda_{2}, \lambda_{3})$-space (Chiueh
\& Lee, 2001).

\begin{figure}
\plotone{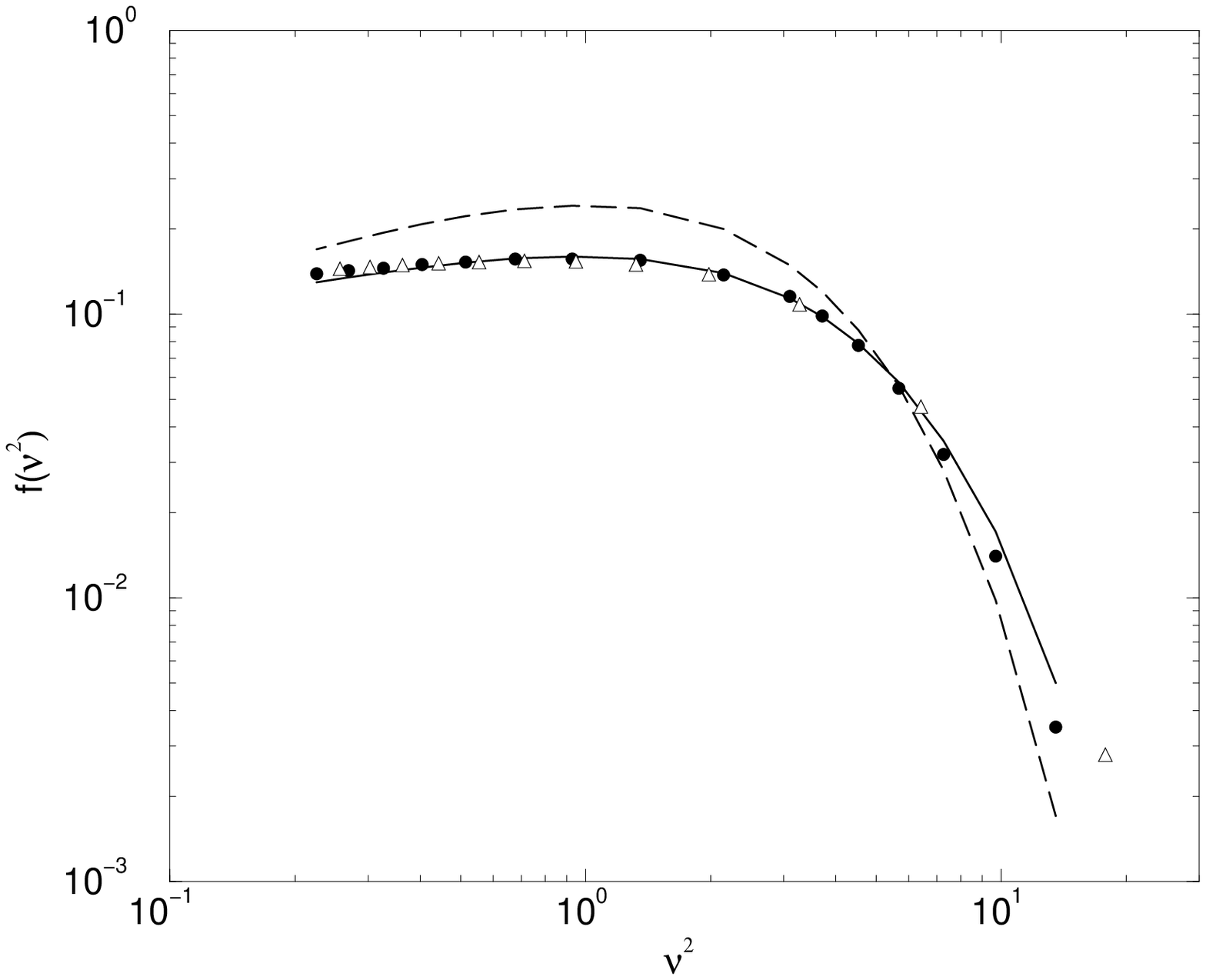} \caption{Comparison of the mass function derived
from EPS model (dashed line) and the nonspherical collapse
boundary (filled circles) with Sheth \& Tormen's fitting formula
of N-body simulations (solid line). The opaque triangles show the
mass function generated from the original nonspherical collapse
boundary proposed by Chiueh \& Lee (2001).} \label{massfun}
\end{figure}

In Fig. \ref{massfun} we compare the mass function derived from
the $Y$-space diffusion with two other types of mass functions.
The filled circles are the data of our six-dimensional random
walks, the dashed line shows the EPS's analytical prediction, and
the solid line shows the Sheth \& Tormen's mass function (ST99),
which has the form
\begin{equation}\label{stmassfun}
\nu f(\nu)=2A[1+(a\nu^{2})^{-q}](\frac{a\nu^{2}}{2\pi})^{1/2}
\mathrm{exp}(-\frac{a\nu^{2}}{2}),
\end{equation}
where $a=0.707$, $q=0.3$ and $A=0.322$. We conclude that our
results agree quite well with that of N-body simulation given by
ST99 (c.f., Eq. (\ref{stmassfun})), except at very high mass
regime. There are two reasons for us to retain this discrepancy at
high mass. First, the statistics of very high mass halos is not
well-controlled in N-body simulation. Therefore one can not
reliably specify the number of such halos. Second, it is believed
that the very high mass halos should follow spherical collapse
process which is closer to the EPS method, and our results show
this tendency. As a further comparison, we also demonstrate in
Fig. \ref{massfun} the mass function given by the original
nonspherical collapse boundary proposed by Chiueh \& Lee (2001).
It is shown that the two NCB mass functions are roughly the same
except that our mass function is more accurate in the intermediate
mass range. Just as we have mentioned earlier, the high accuracy
is necessary when one wants to derive the bias factor precisely
since the bias factor is related to the small difference between
the biased mass function and the unbiased mass function, an issue
to be addressed in the next section.

\subsection{Bias}The peak biasing was originally
introduced by Kaiser (1984) to explain the clustering of Abell
clusters, and the same idea has also been used to discuss whether
the galaxy distribution traces the mass distribution (Bardeen et
al. 1986). Despite focusing on specific bound objects, the bias
parameter also reveals the spatial clustering of dark matter
halos. One common definition of the bias in the Lagrangian
coordinate is
\begin{equation}
\frac{\delta n}{n}=b\frac{\delta \rho}{\rho},
\end{equation}
or
\begin{equation}
b(M)=\frac{n(M,\delta_{c}|\delta=\delta')-n(M,\delta_{c}|\delta=0)}{n(M,\delta_{c}|\delta=0)\times\delta'},
 \label{eq24}
\end{equation}
where $n(M,\delta_{c}|\delta)$ represents the number of bound
objects with mass M which can be found in the large-scale
uncollapsed region of density contrast $\delta$. The second term
of the numerator on the right hand side
$n(M,\delta_{c}|\delta=0)$, or named "the unbiased mass function",
is actually the conventional mass function we have been discussing
so far, and $n(M,\delta_{c}|\delta=\delta')$ is "the biased mass
function" in the sense that it is embedded in the uncollapsed
region with finite density contrast $\delta'$. Mo \& White (1996)
have derived the bias factor under the limit $\delta <<
\delta_{c}$, as shown in Eq. (\ref{eqbias}), and they also give
another equivalent expression for the bias evaluated at a
sufficiently large separation $r$:
\begin{equation}
\xi_{hh}(r,M)=b^{2}(M)\xi_{mm}(r),
\end{equation}
where $\xi(r)$ is the two-point correlation function and the
subscripts $hh$, $mm$ denote the halo-halo and matter-matter
components respectively. This definition is often used to find the
bias factor in N-body simulation.

In the EPS picture, one can calculate $b$ as:
\begin{equation}
b(M)=\frac{n(M,\delta_{c}-\delta'
|\delta=0)-n(M,\delta_{c}|\delta=0)}{n(M,\delta_{c}|\delta=0)\times\delta'}=\frac{1}{\sigma(M)\times
n(\nu)}\frac{dn(\nu)}{d\nu}=\frac{1}{\delta_{c}}\frac{d(\ln
n(\nu))}{d(\ln \nu)},
\end{equation}
indicating that the bias function is directly related to the
derivative of mass function in the log-space. However, this
simplicity is no longer valid in the NCB approach due to the
additional $r$-distribution, and therefore the bias function can
be another test for the goodness of the NCB. The biased mass
function $n(M,\delta_{c}|\delta=\delta')$ can still be obtained
from the excursion set approach. The natural choice for the
initial condition of six-dimensional random walks for the biased
collapse is that $y_{1}=\delta'$, whereas $y_{2}$ to $y_{6}$ are
Gaussian distributed random numbers with the same variance equal
to $\delta'$. Comparing the biased mass function with the unbiased
mass function already obtained in $\S$ 3.1, we are able to compute
the bias function defined in Eq. (\ref{eq24}).

\begin{figure}
\epsscale{0.7} \plotone{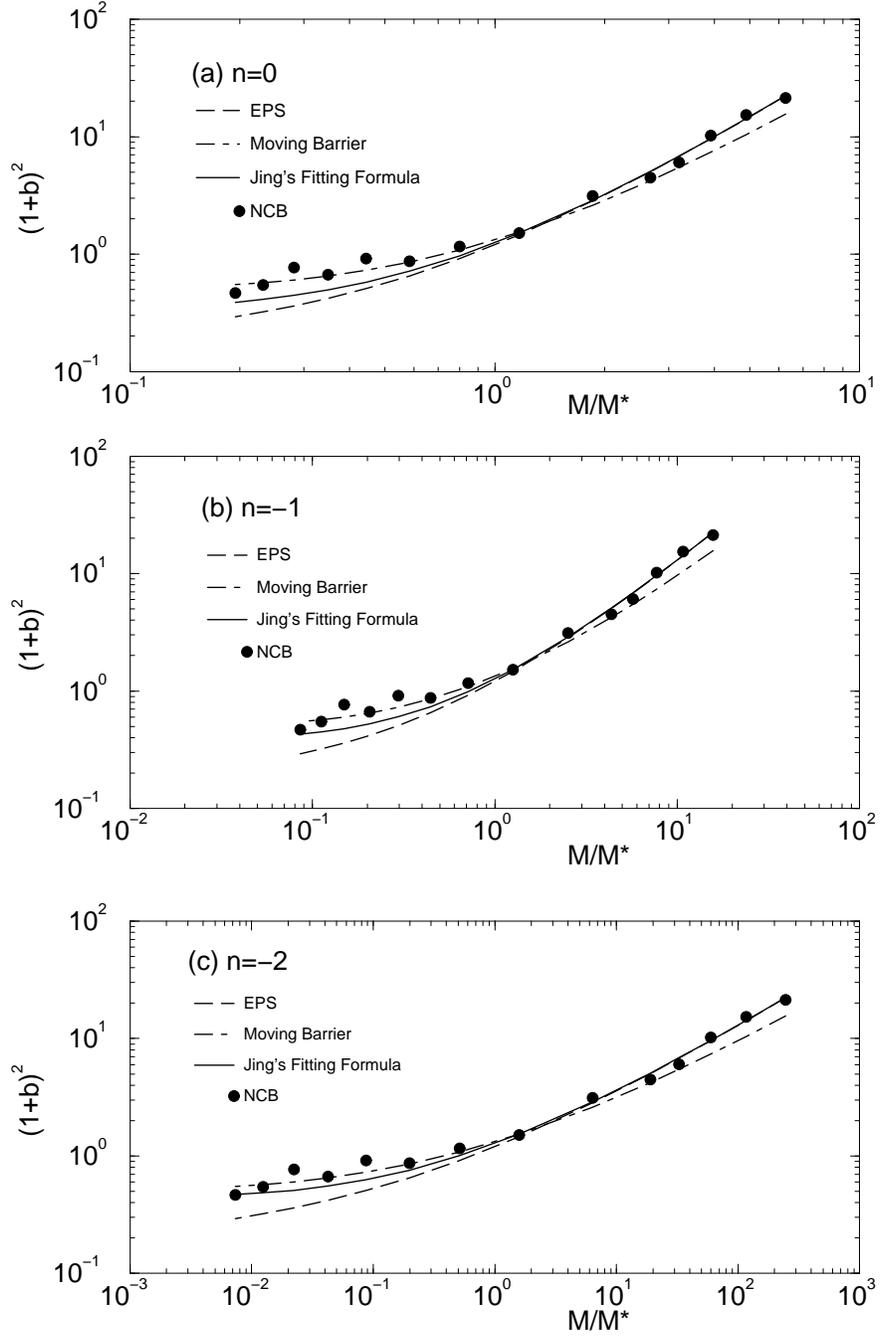}
 \caption{The bias factor as a
function of $M/M^{*}$ for three different initial power
indices:(a) n=0, (b) n=-1 and (c) n=-2. The circle symbols show
the results of our algorithm of $Y$-space random walks. The solid
curves are Jing's fitting formula derived from N-body simulations.
The long dashed lines denote the analytical bias prediction based
on EPS (MW96), and the dot-dashed lines show the results predicted
by moving-barrier model (Sheth, Mo \& Tormen 2001).} \label{bias}
\end{figure}

In Fig. \ref{bias}, we show the bias functions derived from our
approach for three different initial power spectra of indices,
$n=0, -1$, and $-2.$ The filled circles show the results of our
algorithm of $Y$-space random walks. The solid curves denote the
fitting formula of N-body simulation results found by Jing (1998):
\begin{equation}
b(M)=(\frac{0.5}{\nu^{4}}+1)^{(0.06-0.02n)}(1+\frac{\nu^{2}-1}{\delta_{c}})-1.
\end{equation}
As a comparison, Fig. \ref{bias} also shows the predictions of the
EPS model (MW96) and the moving-barrier model (Sheth, Mo \& Tormen
2001). Although there exit statistical fluctuations in our data,
the filled circles show that the bias function generated from the
nonspherical collapse boundary is quite reasonable and has a
similar tendency as the results derived from the moving barrier
model in the low-mass scale. But the curves of our results turn to
coincide with the EPS prediction at high masses. This tendency is
also seen in Jing's simulations.
\section{Conditional Mass Function}

According to the hierarchical scenario, dark matter halos are
formed through merging with comparable-size halos and accretion of
small halos. All kinds of interactions between halos result in
changes of gravitational potential and consequently the behavior
of gas components. Therefore, merger/accretion rates are believed
to play a crucial role in the galaxy formation. Observations show
that merger rates increase with redshift (Carlberg et al. 1994;
Yee \& Ellingson 1995), despite the different definitions of the
merger rate. On the other hand, the dependence of merger rates on
redshifts and on environments has also been studied with N-body
simulations (e.g., Gottl\"{o}ber et al. 2001), and the results
roughly agree with the observations. It is expected that the
semi-analytical approach (e.g., EPS model and NCB model) can offer
succint description for merger histories. However, we must keep in
mind that the semi-analytical methods deal with only the isolated
halos, discarding the information of substructures within each
halo. The attempt of explaining galaxy formation inside galaxy
groups or clusters with semi-analytical approaches must be
modified. Despite such a limitation for the cloud-in-cloud
problem, the merger-tree technique is often needed to establish
merger histories in the field ( Kauffmann \& White 1993;
Somerville \& Kolatt 1999). The merger-tree method is a
Monte-Carlo method based on the conditional mass function. In the
literature, EPS's conditional mass function is often used to
construct merger trees. However, the results indicate that there
still exhibit discrepancies with the merger history extracted from
N-body simulations, which is thought to be caused mainly by the
EPS model rather than the specific features of merger trees
(Somerville et al. 2000). Therefore, in this section, we use the
NCB model to generate a new conditional mass function with the
six-dimensional random walk procedure and derive a fitting formula
for it.
\subsection{Construction of conditional mass function}
\begin{figure}
\plotone{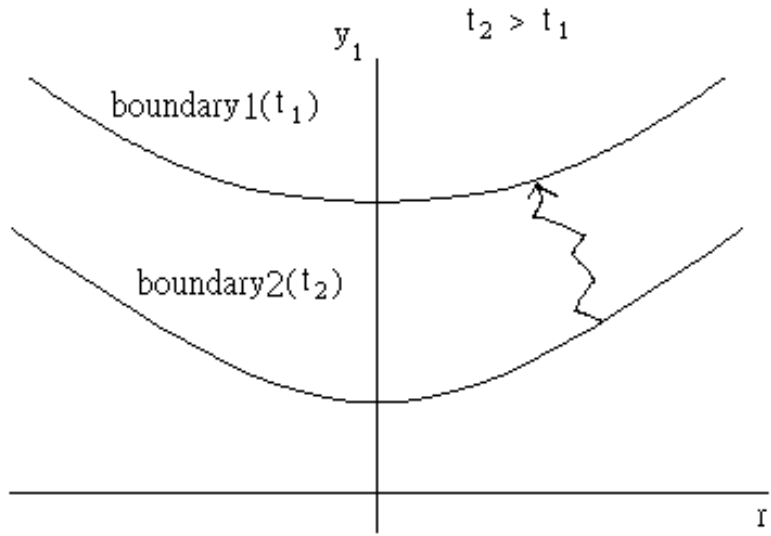} \caption{The nonspherical collapse boundaries of
two different epochs. The higher the boundary is, the earlier the
collapse time.} \label{NCB-merger}
\end{figure}

There are two scenarios, the active picture and the passive
picture, to represent the diffusion process for epochs earlier
than the present. The former is with a fixed collapse boundary and
let the variance of some particular mass grow with time. The
latter is just the opposite which fixes the variance of mass $M$
and lowers the collapse boundary with increasing time. In order to
simulate the merging problem, we adopt the second picture for
simplicity. For a given redshift $z$, the collapse boundary
condition can be rewritten as
\begin{equation}\label{boundary2}
\frac{\delta}{\delta_{c}(z)}=(1+\frac{1}{\beta_{1}}(\frac{r}{\delta_{c}(z)/\delta_{c}(z=0)})^{4})^{\beta_{2}},
(\beta_{1}=0.26, \beta_{2}=0.159)
\end{equation}
where $\delta_{c}(z)$ denotes the vertex of the boundary at
redshift $z$ and $\delta_{c}(z=0)$ is just $\delta_{c}$ given in
Eq. (\ref{boundary}). The evolution of $\delta_{c}(z)$ is
inversely proportional to the growth factor of the variance
$\sigma$. For example, $\delta_{c}(z) \propto (1+z)$ in $S$CDM
model. As shown in the Fig. \ref{NCB-merger}, boundary 1
corresponds to the collapse condition at a time earlier than
boundary 2. That is, Eq. (\ref{boundary2}) is nothing but to say
that boundary 1 should be stretched both in $r$ and $\delta$ axes
relative to boundary 2 to preserve the self-similar evolution of
fluctuations. After collecting the trajectories that start from
the origin and collapse at boundary 2 , we continue the random
walks until they hit boundary1 and are absorbed.

Let $f(S_{1},\delta_{c}(z_{1})|S_{2}, \delta_{c}(z_{2}))$ denote
the fraction of trajectories hitting boundary 1 at mass scale
$S_{1}$ to the total trajectories hitting boundary 2 at mass scale
$S_{2}$. The physical meaning of
$f(S_{1},\delta_{c}(z_{1})|S_{2},\delta_{c}(z_{2}))$ is the
probability that a point in a halo with mass scale $S_{2}$ at
redshift $z_{2}$ was in a halo with mass scale $S_{1}$ at redshift
$z_{1}$. As mentioned earlier, although $S$ is the variance of the
density fluctuation corresponding to a specific mass range, we use
it to interpret the mass scale over which the density contrast
$\delta$ is smoothed. Apparently , $S_{1}$ is always larger than
$S_{2}$ to ensure that the mass of halo increase with time.
Moreover, due to the self-similar scaling property of the
nonspherical collapse boundary, $f$ does not depend on time
explicitly, but only on the ratio of the separation, namely
$\delta_{c}(z_{1})/\delta_{c}(z_{2})\equiv \kappa$. The actual
time evolution of the conditional mass function is hidden in the
variable $\mu$ or $\mu'$ defined below, which is a combination of
mass and time. Therefore, by simulating random walks with a
particular chosen $z_{2}=0$, we are able to construct the
conditional mass function which can also be applied to any other
$z_{2}$.

Recalling the Pess-Schechter's boundary, its conditional
probability is given in Eq. (\ref{conditionaleq}), or in a concise
form,
\begin{equation} \mu f(\mu)d \mu=2\times(\frac{\mu^{2}}{2
\pi})^{1/2}\mathrm{exp}(-\frac{\mu^{2}}{2})d \mu,
\end{equation}
where $\mu
\equiv(\delta_{c}(z_{1})-\delta_{c}(z_{2}))/(S_{1}-S_{2})^{1/2}=\Delta\delta_{c}/(\Delta
S)^{1/2}$. Apparently, for given $z_{1}$ and $z_{2}$, the
conditional probability is a function of only the difference of
the variance $S$ in EPS picture. However, the conditional
probability actually depends also on the initial mass scale
$S_{2}$ in our scenario. This is due to that the separation of two
boundaries is $S_{2}$-dependent, as is obviously indicated in Fig.
\ref{NCB-merger}.

Here we simulate several conditions for different boundary
separations: $\kappa = 1.1$, $1.3$, $2.0$, $3.0$ and $4.0$ with
$z_{2}=0$. It helps the construction of a new conditional mass
function by considering the following two extreme cases: when
$\kappa$ approaches unit (the two boundaries are close), the
conditional mass function is expected to be close to the extend
Press-Schechter's case because the two boundaries are almost
parallel and most points collapsing at the first boundary will
soon collapse in few steps at the second boundary. On the other
hand, when $\kappa$ is very large, points in the second boundary
relative to the first boundary are like that they are located at
the origin, so that the conditional mass function should behave
like the Sheth and Tormen's mass function (hereafter ST mass
function). We find that our simulation data can still be fitted by
the following function:
\begin{equation}\label{fitting}
\mu^{'}f(\mu^{'})d\mu'=2A(\kappa)(1+\frac{1}{\mu^{'^{2q}}
})(\frac{\mu^{'^{2}}}{2\pi})^{1/2}\mathrm{exp}(-\frac{\mu^{'^{2}}}{2})d\mu',
\end{equation}

where
\begin{equation}\label{mu}
\mu^{'}\equiv\frac{[\delta_{c}(z_{1})-\delta_{c}(z_{2})]\varepsilon(S_{2},\kappa)}{(S_{1}-S_{2})^{1/2}},
\end{equation}
\begin{equation}\label{eqA}
A=0.322+\frac{0.178}{\kappa},
\end{equation}
\begin{equation}\label{eqb}
\varepsilon(S_{2},\kappa)=\varepsilon(x)=0.036x^{4}-0.309x^{3}+0.944x^{2}-1.060x+1
\end{equation}
with $x\equiv(\sqrt{S_{2}}/\delta_{c}(z_{2})-0.25)/\kappa$, and
$q(\kappa)$ is required to satisfy the normalization condition
\begin{equation}
A=\sqrt{\frac{\pi}{2}}/(\sqrt{\frac{\pi}{2}}+\frac{\Gamma[-q+\frac{1}{2}]}{2\times\frac{1}{2}
^{(-q+\frac{1}{2})}}).
\end{equation}
Further explanation of the fitting procedure for these fitting
parameters is described in Appendix A.

The mathematical form of Eq. (\ref{fitting}) is similar to the ST
mass function, which demands $A$, $q$ to be constants. Here we
extend the ST mass function to the conditional mass function and
introduce $A(\kappa)$, $q(\kappa)$ and $\varepsilon(S_{2},\kappa)$
as fitting functions. The definition of $\mu^{'}$ here has been so
motivated to fit the probability distribution derived from random
walks.

However, it has been noted that when the time interval is small,
i.e., $z_{2}$ close to $z_{1}$, the concept of random walks fails
to describe the merger events because the excursion set neglects
the correlations between scales (Peacock \& Heavens 1990, Sheth \&
Tormen 2002), and this explains why the conditional mass function
derived from the moving-barrier in a recent report (Sheth \&
Tormen 2002) may not be applied to construct the merger tree.
Although Eqs. (\ref{fitting}) and (\ref{mu}) faithfully describe
the result of our random-walk processes, in order to make the
conditional mass function useful even at small time interval, we
find that $\mu'$ should be modified into
\begin{equation}\label{newmu}
\mu' \equiv
\frac{[\delta_{c}(z_{1})-\delta_{c}(z_{2})]\varepsilon(S_{2},\kappa)}{(S_{1}-S_{2}-\eta\delta_{c}^{2}(z_{2}))^{1/2}}.
\end{equation}
The replacement of $S_{1}(M_{1})-S_{2}(M_{2})$ by
$S_{1}(M_{1})-S_{2}(M_{2})-\eta\delta_{c}^{2}(z_{2})$ means that
there exists a small mass gap between parent halos and progenitor
halos. This small mass gap only becomes important when the
look-back time is small and becomes negligible for a large
look-back time. A small positive $\eta$ serves to correlate the
probability of progenitor halos to that of parent halos. We find
that $\eta=0.06$ gives the best result when compared with
simulations, as discussed below. Furthermore, the variance and
threshold value appear as $\sqrt{S}/\delta_{c}(z_{2})$ in $\mu'$,
such that the active picture and the passive picture are
consistent with each other.

\subsection{Comparison with N-body simulation}In order
to test whether the conditional mass function derived in the
previous section is valid for describing the formation history of
dark halos, we compare our conditional mass function with the
results of the particle-particle/ particle-mesh simulation given
by Somerville et al. (2000). This simulation is a run for the
$\tau$CDM model ($\Omega = 1, \Gamma = 0.21, h=0.5$) with box size
170 Mpc, particle number $256^{3}$. The halos are identified by
using the 'friends-of-friends' (FOF) algorithm.

\begin{figure}
\plotone{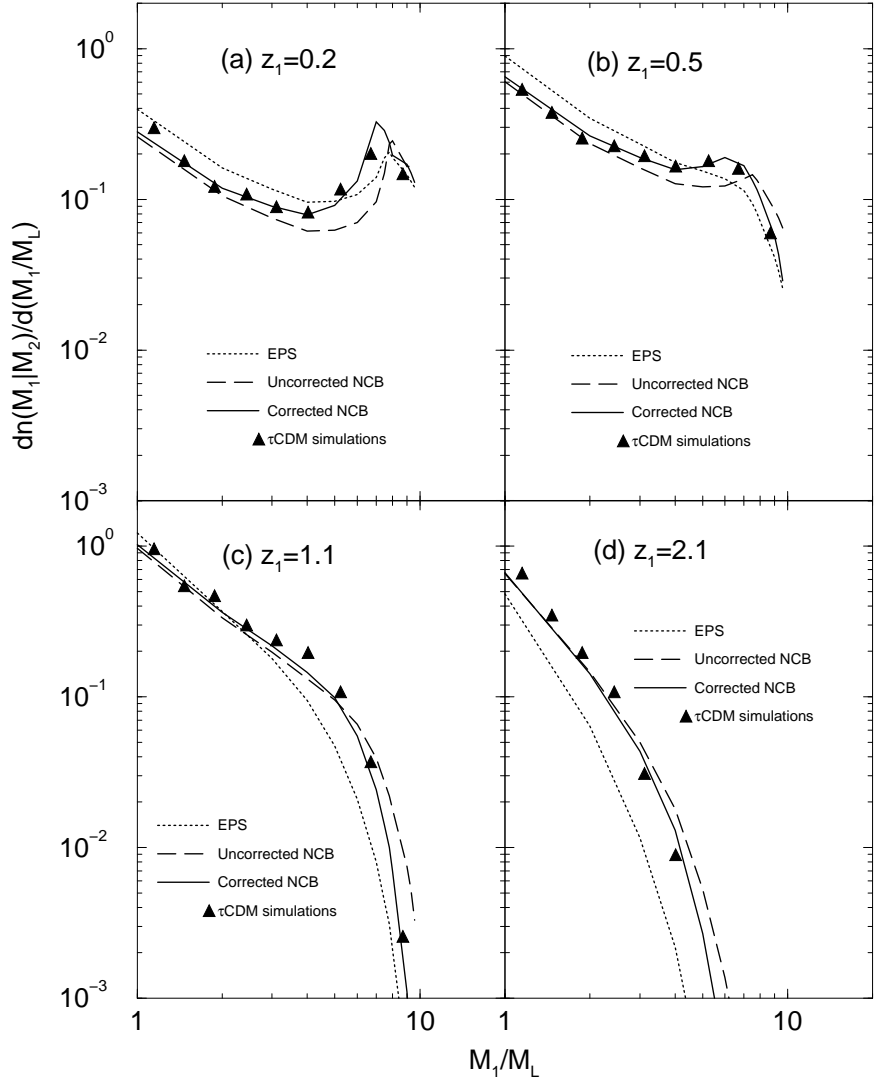}
 \caption{Comparison of
the conditional mass function predicted by EPS (dotted curves),
uncorrected NCB (dashed curves) and corrected NCB (solid curves)
models with the results of the $\tau$CDM simulations (triangle
symbols). The parent halos are chosen at $z=0$, and the
progenitors are at (a) $z=0.2$, (b) $z=0.5$, (c) $z=1.1$ and (d)
$z=2.1$.} \label{newsomer}
\end{figure}

To convert the conditional mass function into the probability that
a halo with mass $M_{2}$ at redshift $z_{2}$ had a progenitor in
the mass range $M_{1}$ to $M_{1}+dM_{1}$ at redshift $z_{1}$, we
have used (LC93)
\begin{equation}
\frac{dn(M_{1},z_{1}|M_{2},z_{2})}{dM_{1}}dM_{1}=\frac{M_{2}}{M_{1}}f(S_{1},\delta_{c}(z_{1})|S_{2},\delta_{c}(z_{2}))|\frac{dS_{1}}{dM_{1}}|dM_{1},
\end{equation}
where
\begin{equation}
 f(S_{1},\delta_{c}(z_{1})|S_{2},\delta_{c}(z_{2})) =
f(\mu')\frac{\partial\mu'}{\partial S_{1}}.
\end{equation}
The mass range of the parent halo chosen by Somerville et al.
(2000) is $7.9 \sim 12.6 M_{L}$ at the present time ($z_{2}=0$),
where $M_{L}\sim 5.0\times 10^{11}M_{\odot}$. Because the parent
mass range is wide, we adopt a fair assumption that the parents of
different mass obey the ST mass function. We use this distribution
of $M_{2}$ to weigh the conditional mass function for the parent
population. In Fig. \ref{newsomer}, we compare the conditional
mass functions predicted by our nonspherical collapse procedure
and by EPS with the results of the $\tau$CDM simulation
(Somerville et al., 2000). The four panels correspond to 4
different redshifts: $z_{1}=0.2, 0.5, 1.1$ and $2.1$. The results
of $\tau$CDM simulation are represented by filled triangles; the
dotted curves denote predictions of the EPS theory; the dashed
curves show predictions by using Eq. (\ref{fitting}) with the
uncorrected definition of $\mu^{'}$ (c.f., Eq. (\ref{mu})). The
solid curves, which are closer to the simulation results are our
predictions with the corrected definition of $\mu^{'}$ (c.f., Eq.
(\ref{newmu})). It can be seen that introducing the small
correction $\eta$ in Eq. (\ref{newmu}) helps reproduce the
conditional mass function more accurately for massive progenitor
halos at small $z_{1}$. The solid curves and dashed curves can
almost coincide when $z_{1}$ gets larger, as the effect of $\eta$
becomes negligible. The comparison shows that our conditional mass
function (c.f., Eq. (\ref{fitting}) \& Eq. (\ref{newmu})) provides
a better description than EPS does for a wide range of redshift.
The successful description of our conditional mass function
suggests that the nonspherical collapse boundary can be more
"realistic" and may be applied to construction of merger trees.

\section{Conclusion}

In this paper, we have attempted to show that the nonspherical
collapse boundary (NCB) reproduces the mass function, bias factor
and conditional mass function better than the Press-Schechter
theory. This model successfully addresses these subjects over a
wide range of mass scales, while the (extend) Press-Schechter
theory always fails in some mass range.

Through the simulations of 6-dimensional random walks, we have
shown that the bias function generated by NCB agrees with the
N-body simulations reasonably well. Because NCB is more
complicated than the PS's collapse boundary, the relation between
the bias function and the conditional mass function is no longer
as simple as the EPS model. Especially, the two issues are not of
identical representation in the framework of NCB model. We provide
a formula for the conditional mass function of the nonspherical
collapse boundary, which can be applied to any two redshifts.  In
order to overcome the problem that random-walks fail to describe
the merger history at small look-back time, we add a minor
correction in the definition of variable $\mu'$ (c.f. Eq.
(\ref{newmu})) while maintaining the overall mathematical form
intact. This formula can describe the conditional mass probability
more accurately than the EPS formalism over a wide mass range,
even when the two epochs are close. All the results we obtained
from NCB suggest that NCB is sufficiently reasonable to reproduce
the statistical properties of dark halos.

There are two interesting issues pertinent to NCB: one is its
natural association with the halo angular momentum through its
dependence on the quantity $r$. The halo angular momentum
distribution predicted by the NCB model is reported elsewhere
(Chiueh, Lee and Lin, in preparation). Second, as mentioned in
$\S$ 4, the conditional mass function offers an useful tool to
construct the merger history through merger trees. Recent works
have indicated that the merger history constructed by using EPS's
conditional mass function deviates from that extracted from N-body
simulations, and it is thought to be caused mainly by the
spherical-symmetry assumption of the EPS model rather than the
specific features of the merger tree scheme. Thus, we expect the
agreement can be greatly improved by the NCB conditional mass
function given in this work.

\acknowledgments We thank Dr. Y. Jing for useful comments on this
paper. We also thank Dr. H. Mo and Dr. R. Wechsler for pointing
out typos. This work is supported in part by the National Science
Council of Taiwan under the grant, NSC-90-2112-M-002-026.

\appendix

\section{The fitting procedure of conditional mass function}
\begin{figure}
 \plotone{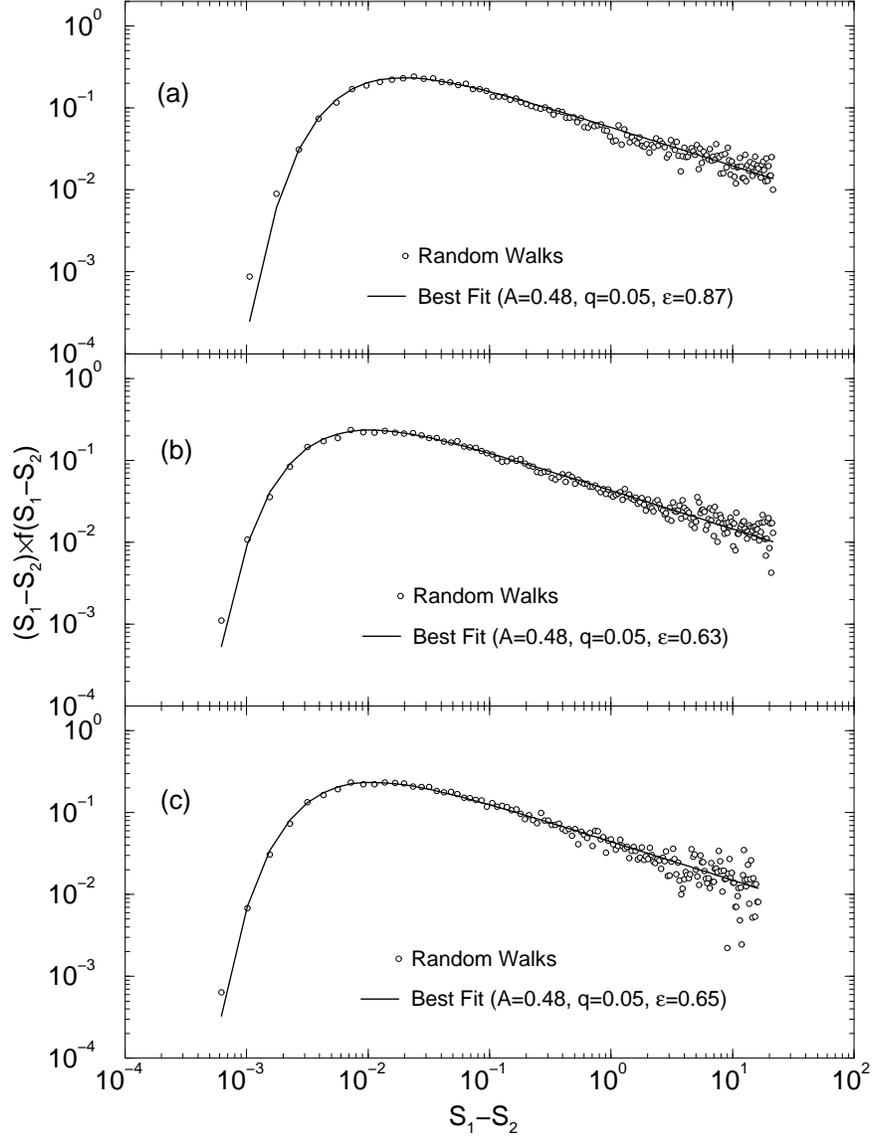}
\caption{The conditionalmass function for $\kappa=1.1$. The opaque
circles represent the results of the six-dimensional random walks
and the solid lines are the best fit using Eqs. (\ref{fitting})
and (\ref{mu}). The mass scale at time $z_{2}$ is chosen as (a)
$\sqrt{S_{2}}=0.7$, (b) $\sqrt{S_{2}}=1.5$ and (c)
$\sqrt{S_{2}}=2.9$.} \label{i8}
\end{figure}

\begin{figure}
 \plotone{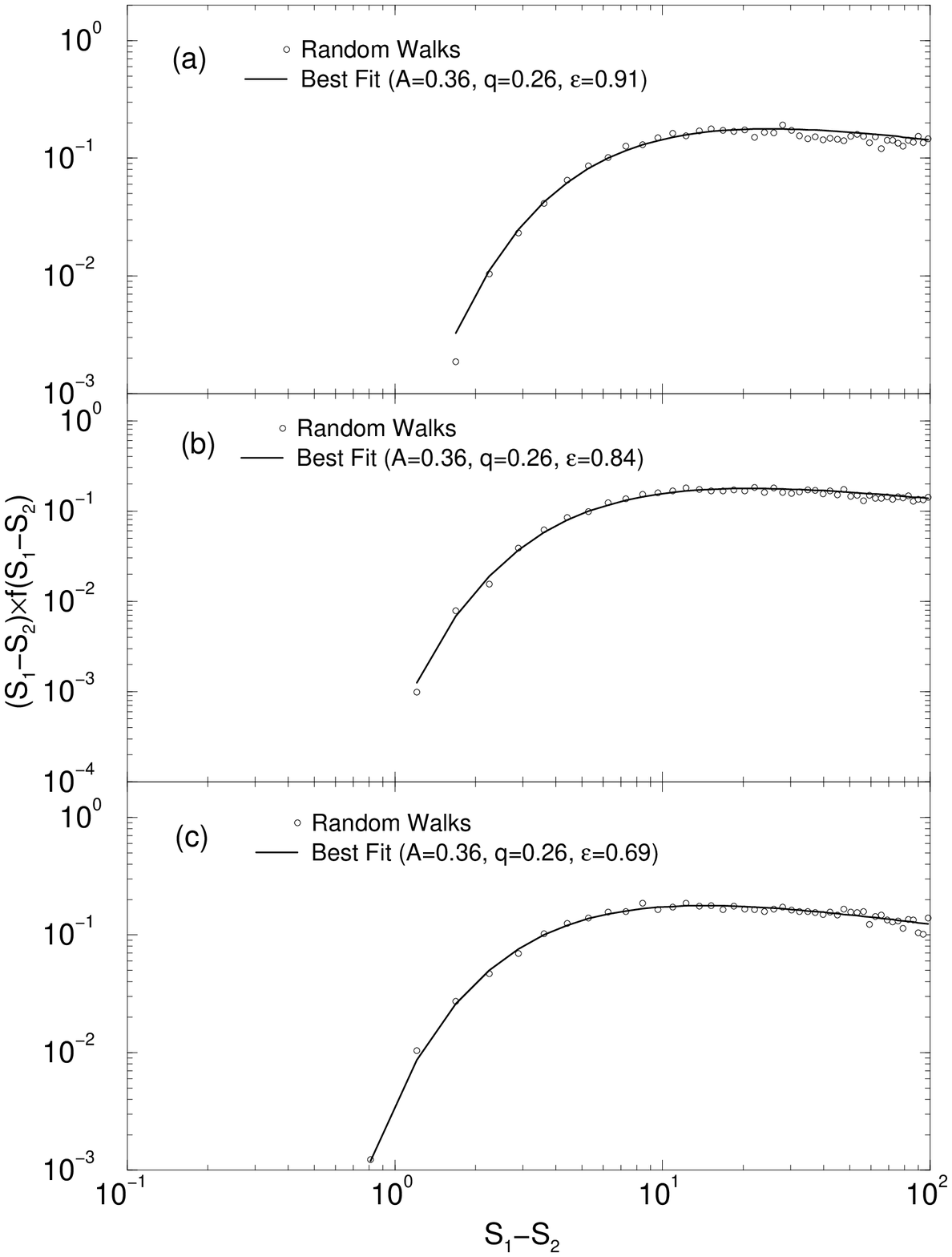}
\caption{Same as the previous figure, but for $\kappa=4.0$.}
\label{k8}
\end{figure}

After performing the random walks between two collapse boundaries,
which are mentioned in $\S$ 4.1, we use the mathematical form of
ST mass function and tune the values of A, q and $\varepsilon$ to
adjust our results of random walks. It should be emphasized that A
and q are not tuned independently because the integral of the
function $f(\mu^{'})$ should be normalized to unity, due to the
assumption that all the trajectories starting from the first
boundary will eventually hit the second boundary. Thus, we have
\begin{equation}
\int f(\mu^{'})d\mu^{'}=\int 2A(\kappa)(1+\frac{1}{\mu^{'^{2q}}
})(\frac{1}{2\pi})^{1/2}\mathrm{exp}(-\frac{\mu^{'^{2}}}{2})d\mu'=1,
\end{equation}
then we get
\begin{equation}\label{aq}
A=\sqrt{\frac{\pi}{2}}/(\sqrt{\frac{\pi}{2}}+\frac{\Gamma[-q+\frac{1}{2}]}{2\times\frac{1}{2}
^{(-q+\frac{1}{2})}}).
\end{equation}
In addition, it is expected that all three parameters $A$, $q$ and
$\varepsilon$ should yield a fitting formula of conditional
probability to match the two extreme cases, as mentioned in $\S$
4.1. For $\kappa$ approaching unit, i.e, the two boundaries are
very close, $A(\kappa)$ gives the value $\frac{1}{2}$, and q is
thus zero, which agrees with the Press-Schechter's conditional
probability. For $\kappa$ to be very large, $A$ and $q$ approach
0.322 and 0.3 respectively, the values given by Sheth \& Tormen
(1999). The results of our 6-dimensional simulation and our
fitting formula Eq. (\ref{fitting}) and (\ref{mu}) are shown in
Fig. \ref{i8} $\sim$ Fig. \ref{k8}. The five figures correspond to
different separations of boundaries, i.e., $\kappa =
1.1,1.3,2.0,3.0$, and $4.0$. The top panel of each figure is the
result for initial mass scale $\sqrt{S_{2}}=0.7$, the middle panel
is for $\sqrt{S_{2}}=1.5$ (around $M^{*}$) and the bottom is for
$\sqrt{S_{2}}=2.9$. The opaque circles represent the data from
6-dimensional simulation and the solid lines are the best fit
using Eq. (\ref{fitting}) and (\ref{mu}).

\begin{figure}
\plotone{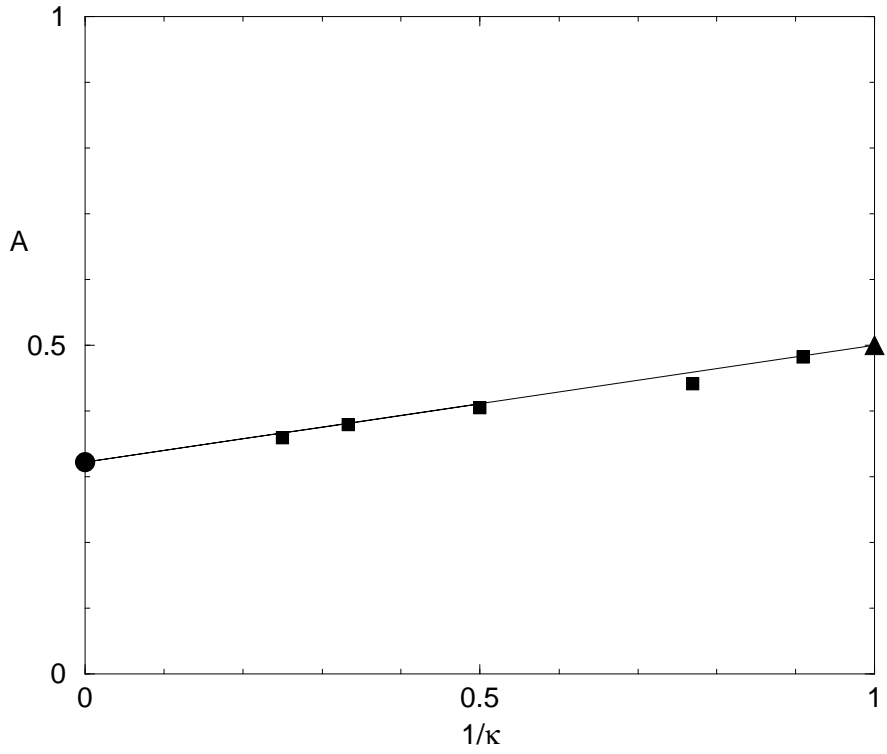} \caption{$A-1/\kappa$ relation. The square
symbols show the five fitting values of $A$ for our data, the
filled triangle represents the EPS case, the filled circle
represents the case of ST mass function, and the solid line
denotes the linear relation between $A$ and $1/\kappa$ (c.f., Eq.
(\ref{eqA})).} \label{A}
\end{figure}

We find that for the case $\kappa$ is fixed, $A$ does not vary
with the initial mass scale $S_{2}$, i.e., $A$ is a function of
$\kappa$ only. In Fig. \ref{A}, we show the relation between $A$
and $1/\kappa$.  The square symbols show the fitting values of A
for five different $\kappa$ plus two boundary values:
$(1/\kappa,A)=(0,0.322)$ and $(1,0.5)$, which recover the limit
conditions associated with EPS and ST cases. The solid line
represents here is a linear relation between $A$ and $1/\kappa$:
\begin{equation}
A=0.322+\frac{0.178}{\kappa},
\end{equation}
which is determined by the above two boundary points. This line
almost passes through the other five fitting points and thus is a
good representation for $A(\kappa)$. Combined with Eq. (\ref{aq}),
$q$ can be determined immediately for each $\kappa$.

\begin{figure}
\plotone{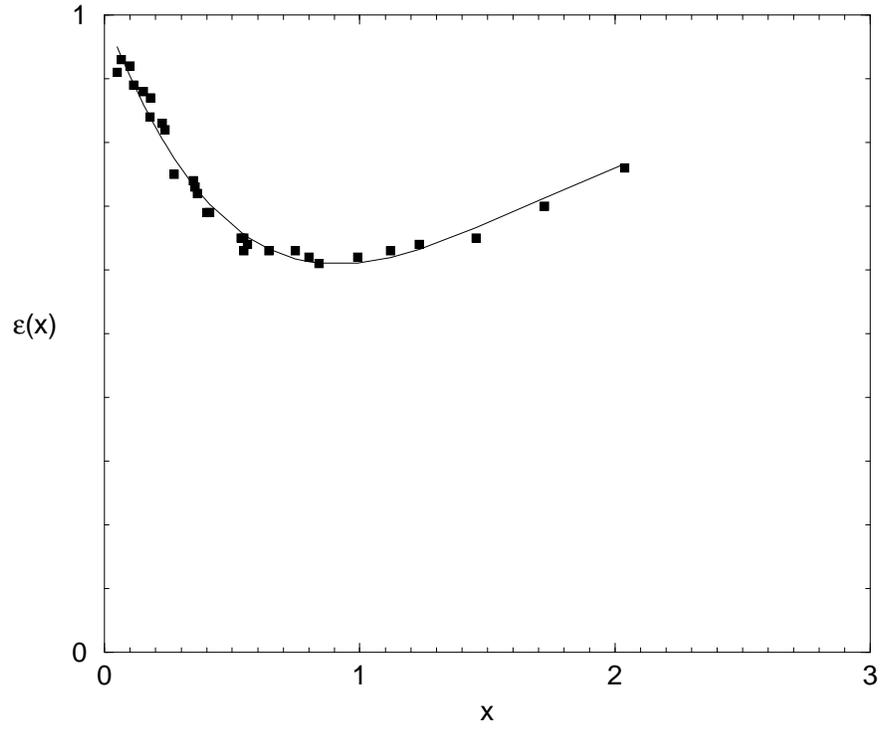} \caption{$\varepsilon$ factor as a function of
$x (\equiv(\sqrt{S_{2}}/\delta_{c}(z_{2})-0.25)/\kappa)$. Similar
to Fig. \ref{A}, the square symbols show the fitting values of
$\varepsilon$ for our data and the solid line denotes the fitting
relation of the square symbols (c.f., Eq. (\ref{eqb})).} \label{b}
\end{figure}

Our tests also show that $\varepsilon$ is a function of a combined
variable $x\equiv(\sqrt{S_{2}}/\delta_{c}(z_{2})-0.25)/\kappa$.
The best fit of the function is (see Fig. \ref{b})
\begin{equation}
\varepsilon(x)=0.036x^{4}-0.309x^{3}+0.944x^{2}-1.060x+1.
\end{equation}
This result is consistent with the geometry of two boundaries.
Since we have the $r$-distribution of various mass scales
$S_{2}$'s that hit the first boundary, one can estimate roughly
the distance between two boundaries for each mass scale. It is
easily seen that for a fixed $\kappa$, the gap $d$ initially
decreases to reach a minimum and then rises when $S_{2}$ continues
to increase. The behavior of
$d/(\delta_{c}(z_{1})-\delta_{c}(z_{2}))$ agrees with that of
$\varepsilon$ defined above.

\clearpage

\clearpage
\clearpage
\clearpage
\clearpage
\clearpage
\begin{figure}
 \plotone{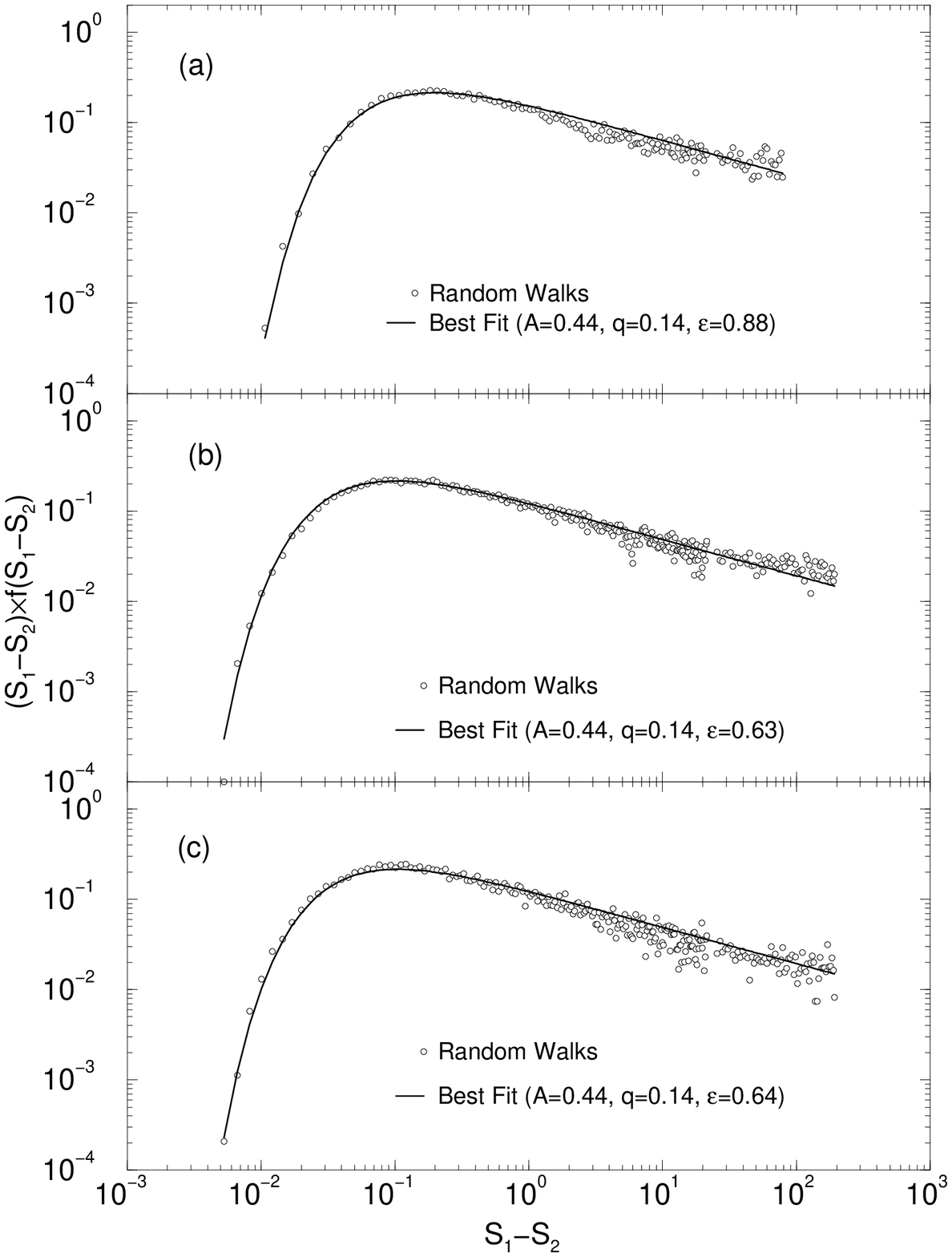}
\caption{Same as the previous figure, but for $\kappa=1.3$.}
\label{e8}
\end{figure}
\clearpage
\begin{figure}
 \plotone{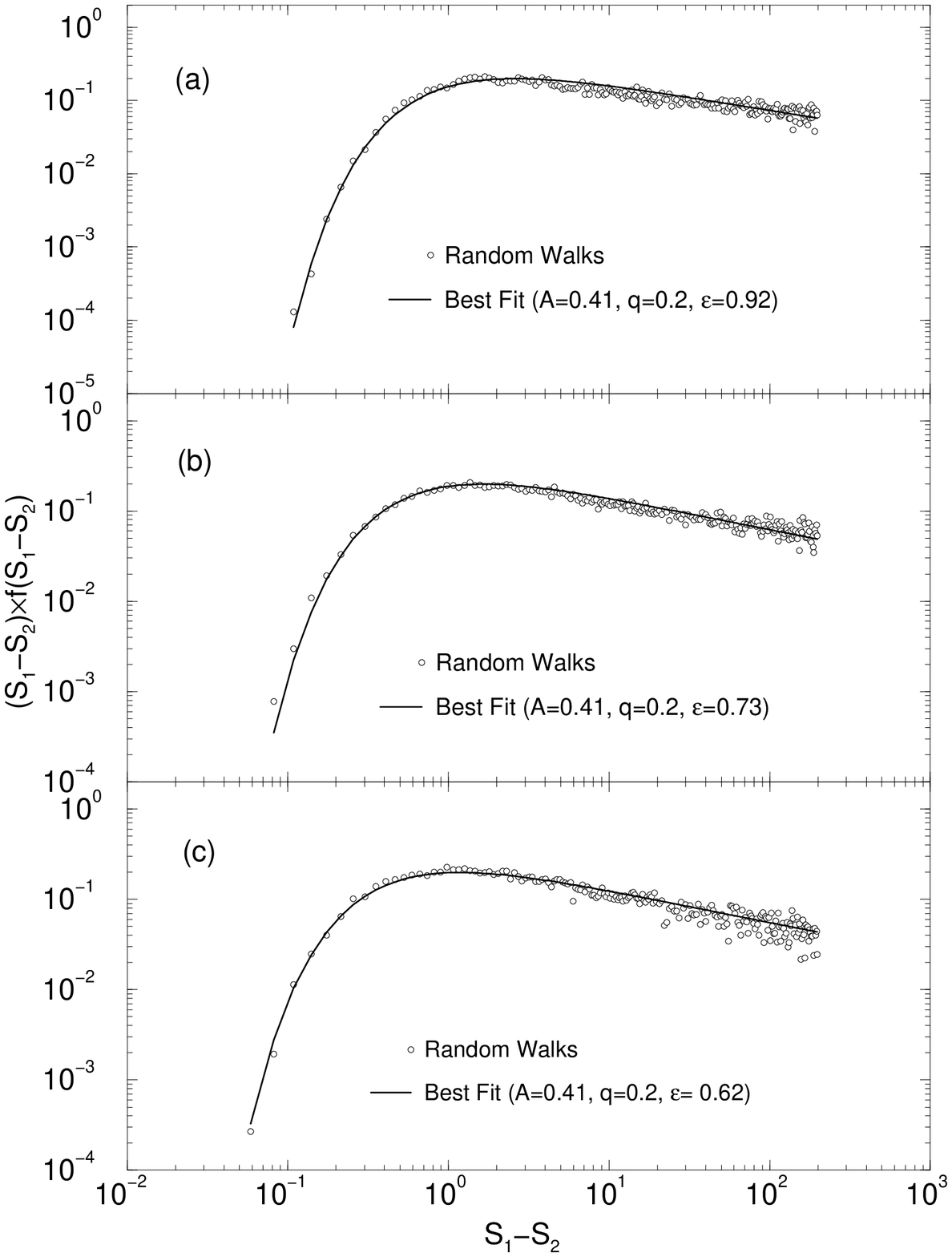}
\caption{Same as the previous figure, but for $\kappa=2.0$.}
\label{f8}
\end{figure}
\clearpage
\begin{figure}
 \plotone{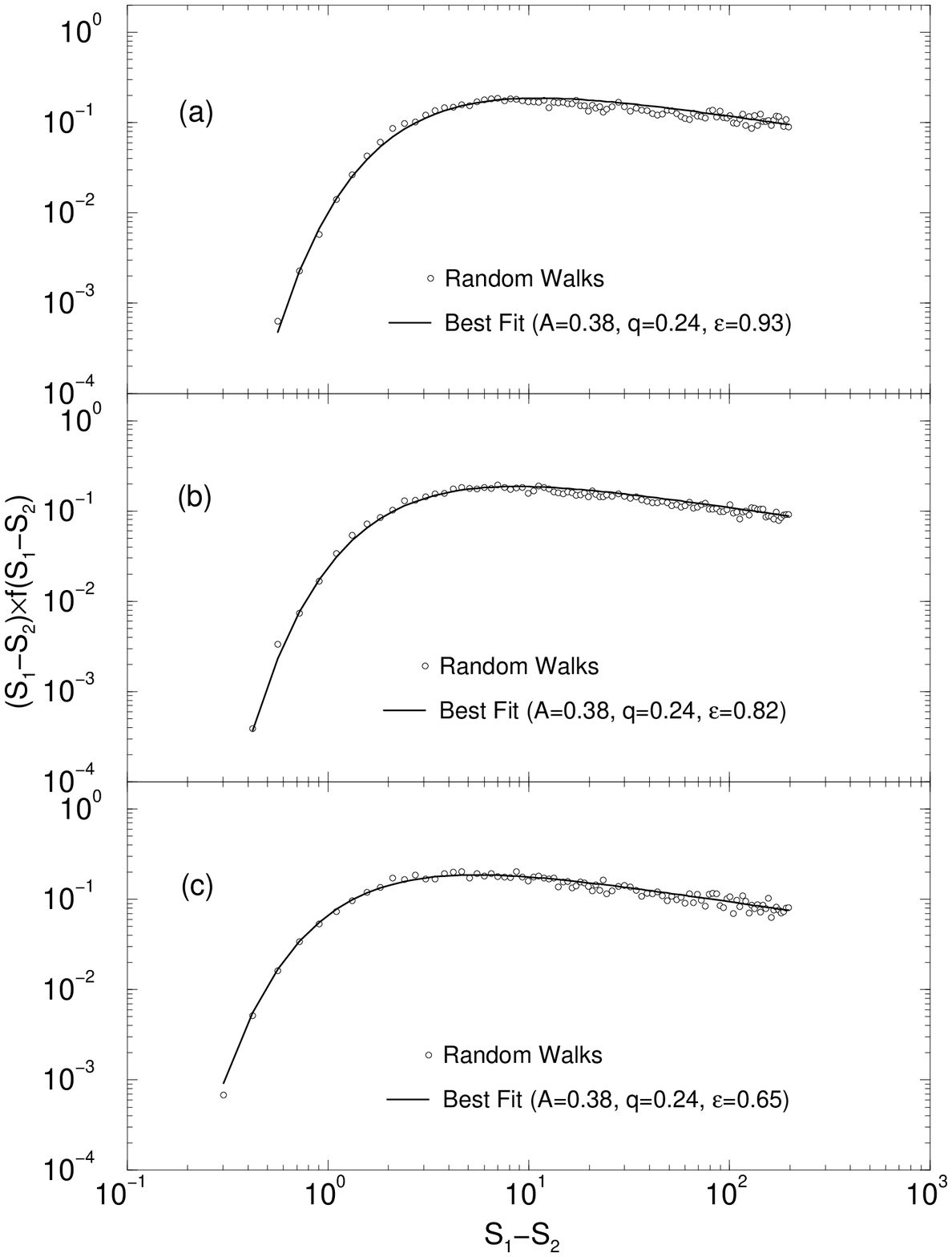}
\caption{Same as the previous figure, but for $\kappa=3.0$.}
\label{g8}
\end{figure}
\clearpage
\clearpage
\clearpage

\end{document}